\documentclass[10pt,a4paper,twoside]{article}
\usepackage{epsfig}
\usepackage{baltlat6}
\usepackage{array}
\usepackage{rotating}       
\usepackage{graphicx}
\usepackage{amsmath} 
\usepackage{amssymb} 
\usepackage{natbib}
\begin{document}

\vspace{0.5mm}
\setcounter{page}{1}
\vspace{8mm}

\titlehead{Baltic Astronomy, vol.\,XX, XXX--XXX, 2011}

\titleb{MEASURING THE COSMIC WEB}

\begin{authorl}
\authorb{Volker M\"uller}{1}, 
\authorb{Kai Hoffmann}{2}, and
\authorb{Sebasti\'an E. Nuza}{3}
\end{authorl}

\begin{addressl}
\addressb{}{Leibniz-Institut f\"ur Astrophysik Potsdam (AIP)\\
An der Sternwarte 16, D-14482 Potsdam, Germany}

\addressb{1}{vmueller@aip.de}
\addressb{2}{hoffmann@ice.cat}
\addressb{3}{snuza@aip.de}

\end{addressl}

\submitb{Received: 2011 July 10; accepted: 2011 XXX XX}

\begin{summary} 
A quantitative study of the clustering properties of the cosmic web as a function of absolute 
magnitude and colour is presented using the SDSS Data Release 7 galaxy survey. Mark correlations 
are included in the analysis. We compare our results with mock galaxy samples obtained with four 
different semi-analytical models of galaxy formation imposed on the merger trees of the Millenium 
simulation. The clustering of both red and blue galaxies is studied separately.
\end{summary}

\begin{keywords} Cosmology:  observations -- cosmology: theory -- galaxies: statistics 
-- large scale structure of the Universe \end{keywords}

\resthead{Measuring the Cosmic Web}
{V. M\"uller, K. Hoffmann, S. E. Nuza}

\sectionb{1}{INTRODUCTION}

The 200 years history of the Tartu Observatory is strongly linked with the exploration of the Earth 
and space at different scales. In the early 19th century, the triangulation along the Tartu Meridian 
Arc, 3000 km across Europe, helped to determine the size and precise shape of the Earth. First 
stellar parallax measurements (besides Bessel) by Wilhelm Struve, the founder of the Tartu Observatory, 
provided the basis for exploring our neighborhood within the Milky Way. The dynamical distance 
measurements of the Andromeda nebula and other {\it island universes} by Ernst \"Opik in 1918--1922 
opened the way to the first systematic works in the field of extragalactic astronomy. 

The study of the large scale distribution of galaxies became an important research subject already 
over 50 years ago with the notion of filamentary structure as revealed by the Lick galaxy survey 
(Shane \& Wirtanen 1954). The impression of a cellular structure of the Universe with dominance of 
filaments and large voids in the galaxy distribution was developed during the period 1974-1980 at the 
cosmology school of Tartu Observatory (Joeveer, Einasto \& Tago 1978; Einasto, Joeveer \& Saar 1980). 
These results were presented at the IAU Symposium No. 79 at Tallinn (Longair \& Einasto 1978) where 
an exposition of the pancake theory of large scale structure formation was presented by Zel'dovich, 
Doroshkevich, Shandarin, Sigov and Kotok  (see e.g. Zel'dovich 1978). Already at this time, galaxy 
formation in proto-clusters was discussed by Doroshkevich, Saar \& Shandarin (1978). 

A quantitative description of the galaxy clustering was provided for the first time by Totsuji \& 
Kihara (1969) establishing the power law dependence of the angular auto-correlation function. However, 
the true spatial distribution became obvious only with the advent of the Harvard-Smithonian Center for 
Astrophysics redshift surveys (Huchra et al. 1983; Geller \& Huchra 1989). The quantitative properties 
of the spatial clustering were provided by Davis \& Peebles (1983) and Efstathiou \& Jedrzejewski (1984). 
Later, more extended surveys confirmed the power law behaviour of the correlation function, in particular 
the Automatic Plate Measuring survey (Efstathiou 1993); the Las Campanas Redshift Survey (Tucker et al. 
1997); the Two-degree-Field Galaxy Redshift Survey (Madgwick et al. 2003), and the Sloan Digital Sky Survey 
(Li et al. 2006, Swanson et al. 2008). In these and related studies it was shown that the clustering of 
galaxies strongly depends on their magnitudes, morphological types, and colours (e.g. Davis \& Geller 1976; 
Loveday et al. 1995; Zehavi et al. 2010). 

We have been involved in a detailed analysis of the cosmic web using both modern redshift surveys and 
numerical simulations of galaxy formation together with colleagues from Tartu. Building on standard 
techniques such as those used in Tucker et al. (1997) we analyze here the largest SDSS galaxy redshift 
catalogue presently available. We also present an analysis of mark correlation functions. The aim of this 
contribution is to investigate the distribution of galaxies and its relation to the underlying dark matter 
density field within the standard $\Lambda$CDM paradigm. We perform a correlation analysis depending on 
the absolute magnitude and colour of observed galaxies and compare the results with a series of 
semi-analytical models of galaxy formation imposed on the Millenium simulation (Springel et al. 2005).

\sectionb{2}{DATA AND MOCK SAMPLE SELECTION}

We study the cosmic web using the SDSS Data Release 7, the largest near field galaxy redshift survey 
available. The survey is complete and comprises a large contiguous region of the Northern Galactic cap 
with 7500 deg$^2$. Photometric calibration and $k$-correction to redshift $z=0$ is done according to 
Hogg et al. (2002) using the galactic extinction measurements of Schlegel et al. (1998). We employ 
absolute Petrosian (1976) AB-magnitudes and use the New York University Value-Added Galaxy Catalog 
(Blanton et al. 2005). 

Starting from the observed $R$-band magnitude and redshift distributions, we define two sets of volume-limited 
galaxy samples as illustrated in Fig. 1 (see Table 1). The first set of volume-limited samples (m1 to m12) 
is used to investigate the dependence of the auto-correlation function on absolute magnitude. The samples
are selected in order to cover a large magnitude range and to enclose a sufficient number of galaxies for
the analysis. Therefore, the samples partially overlap, each separate sample contains however a significant
number of independent objects to derive the auto-correlation functions. The second set (r1, r2, r3) was
selected to cover a large range of magnitudes. This allows us to investigate the magnitude dependence of
clustering using mark correlation functions. We impose a subdivision into red and blue galaxies applying
least squares fitting through the green valley in the $U-R$ and $R$ plane, which leads to a separation line
$U - R = 1.8 - 0.05\times (R+19)$.

For comparison we use four sets of mock galaxy samples constructed using the Millenium simulation. 
It follows the evolution of dark matter haloes and sub-haloes using $2160^3$ particles 
in a large box of 500 $h^{-1}$ Mpc length on a side. Galaxy catalogues are modeled using semi-analytical 
models of galaxy formation from merger trees of haloes in the simulation. The model of 
Croton et al. (2006, hereafter C06) implements AGN feedback in two channels to efficiently suppress star 
formation in high mass haloes (`quasar' and `radio' modes), thereby forming a realistic population of 
elliptical galaxies. The model of De Lucia \& Blaizot (2007, hereafter D07) builds on the first model and 
improves the treatment of satellite mergers, using a more realistic dust model and a different initial 
mass function for the stellar population synthesis. The third catalogue of mock galaxies, produced by 
Font et al. (2008), includes a modelling of ram pressure stripping of satellite galaxies by hot gas inside 
large dark matter haloes. In this way, the luminosity function of faint red galaxies is better reproduced. 
Finally, the model of Guo et al. (2011, hereafter G11) improves the treatment of the cooling flow regime 
and the rapid gas inflow, and it updates some parameters related with star formation and feedback processes. 
The mock galaxy samples are constructed applying the same angular selection as in the observations as well as 
the magnitude and redshift ranges provided in Table 1.

\begin{table}[!t]
\begin{center}
\vbox{\footnotesize\tabcolsep=3pt
\parbox[c]{124mm}{\baselineskip=10pt
{\smallbf\ \ Table 1.}{\small\
Properties of the SDSS volume-limited samples. The correlation length, $r_0$, of the different samples is 
given for samples m1 -- m12 (for blue galaxies only m1 -- m7).\lstrut}}
\begin{tabular}{c|ll|ll|rrr|lll}
 \hline
 Sample & $R_{\rm low}$ & $R_{\rm up}$ & $z_{\rm low}$ & $z_{\rm up}$  & Number & Red & Blue & $r_0(\rm all)$ 
        &$r_0(\rm red)$ &$r_0(\rm blue)$ \hstrut\lstrut\\
\hline
m1  & $-18.35$ & $-19.86$ & 0.020 & 0.056 &  42\,165 &  17\,801 & 24\,364 &  6.33 &  8.72 & 4.58 \hstrut \\
m2  & $-19.08$ & $-20.43$ & 0.026 & 0.078 &  86\,272 &  45\,531 & 40\,741 &  6.45 &  7.83 & 4.81 \\ 
m3  & $-19.73$ & $-20.94$ & 0.032 & 0.105 & 129\,802 &  79\,097 & 50\,705 &  7.29 &  8.35 & 5.36 \\
m4  & $-20.28$ & $-21.40$ & 0.040 & 0.136 & 161\,913 & 107\,837 & 54\,076 &  7.49 &  8.26 & 5.65 \\
m5  & $-20.76$ & $-21.82$ & 0.049 & 0.169 & 161\,392 & 114\,573 & 46\,819 &  8.22 &  8.85 & 6.16 \\
m6  & $-21.16$ & $-22.20$ & 0.058 & 0.20  & 172\,264 &  94\,975 & 32\,289 &  8.94 &  9.74 & 6.99 \\
m7  & $-21.49$ & $-22.54$ & 0.068 & 0.20  &  69\,787 &  55\,468 & 14\,419 &  9.07 &  9.60 & 7.70 \\
m8  & $-21.77$ & $-22.86$ & 0.078 & 0.20  &  32\,677 &  27\,432 &  5\,245 & 10.11 & 10.50 & \\
m9  & $-21.98$ & $-23.16$ & 0.090 & 0.20  &  15\,545 &  13\,597 &  1\,948 & 11.40 & 11.79 & \\
m10 & $-22.15$ & $-23.43$ & 0.102 & 0.20  &   8\,343 &   7\,483 &    860 & 12.07 & 12.45 & \\
m11 & $-22.26$ & $-23.70$ & 0.116 & 0.20  &   5\,077 &   4\,614 &    463 & 12.81 & 13.05 & \\
m12 & $-22.36$ & $-23.96$ & 0.130 & 0.20  &   3\,120 &   2\,856 &    264 & 13.29 & 13.70 & \\
r1  & $-18.51$ & $-20.77$ & 0.03  & 0.06  &  63\,546 &  31\,464 & 32\,082 &       &       & \\
r2  & $-19.39$ & $-22.28$ & 0.06  & 0.09  & 125\,491 &  76\,733 & 48\,758 &       &       & \\
r3  & $-20.01$ & $-23.16$ & 0.09  & 0.12  & 114\,266 &  74\,612 & 39\,654 &       &       & \lstrut \\
\hline
\end{tabular}
}
\end{center}
\vskip3mm
\end{table}

\begin{figure}[!tH]
\vbox{
\centerline{\psfig{figure=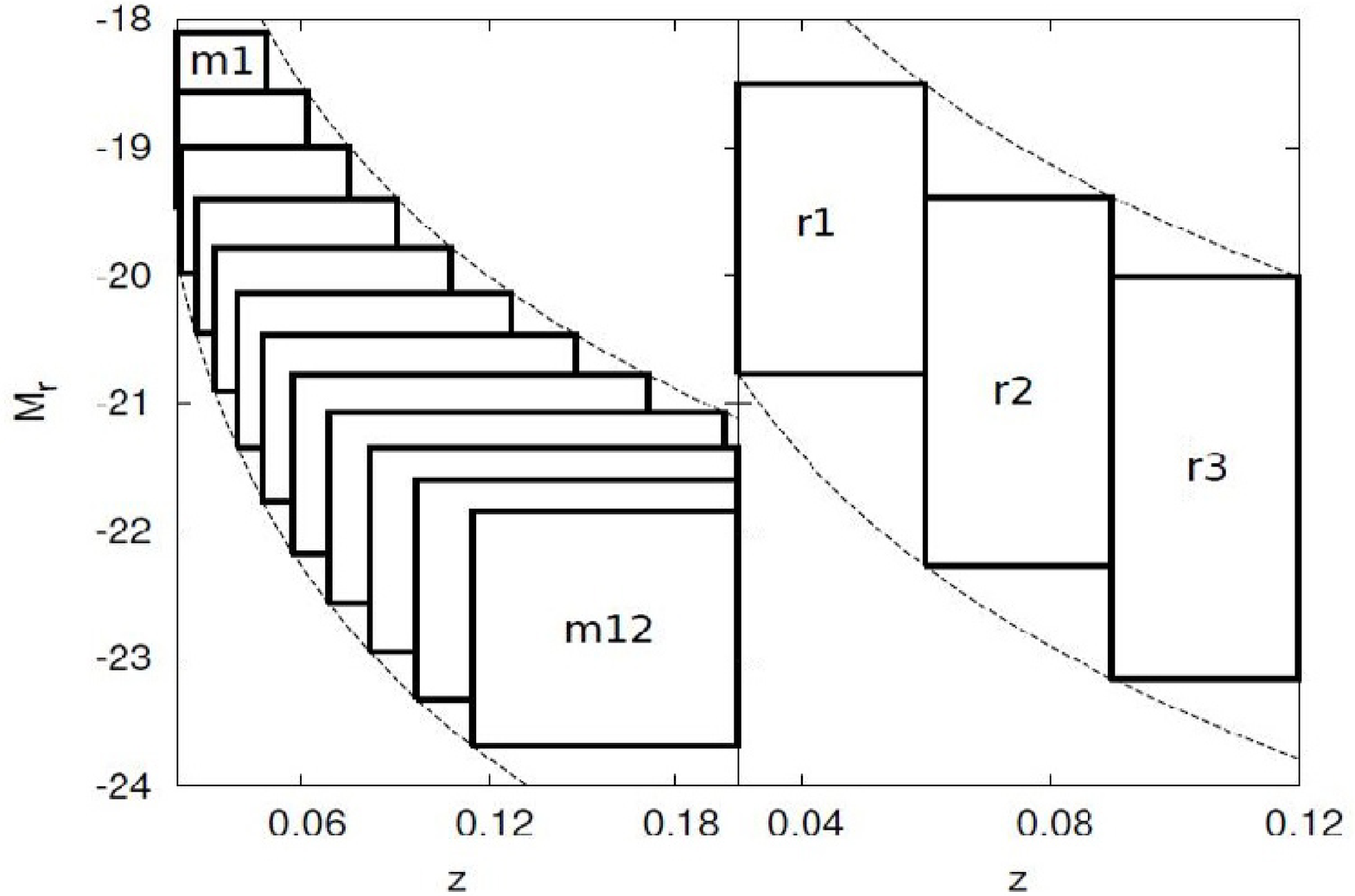,width=60mm,angle=270,clip=}}
\vspace{1mm}
\captionb{1}
{Magnitude and redshift boundaries of the 12 (left panel) and 3 (right panel) 
volume-limited galaxy samples for a large coverage in depth ($m$ samples) 
and magnitude ($r$ samples), respectively.}
}
\end{figure}

\sectionb{3}{CORRELATION ANALYSIS}

The correlation functions are evaluated using the Landy \& Szalay (1993) estimator. 
Data-data, data-random and random-random pairs are generated with the same angular 
selection function of observations and the redshift bounds given in Table 1, however 
not taking into account the fiber separation limit of the SDSS. The 
estimator reads as follows

$$ \xi(r) = \frac{\langle DD(r) - 2 DR(r) + RR(r)\rangle}{\langle RR(r) \rangle}.$$

\noindent Errors are estimated using 10 bootstrap resamplings of the data. Fig. 2 shows 
the convex form of the correlation function over the range from $0.2-50$ $h^{-1}$ Mpc. 
The solid line in the left panel shows the result corresponding to all galaxies for the 
sample m1. Additionally, a power law fit at the correlation length scale, i.e. where 
$\xi(r)=1$, is also shown. The dashed line stems from red galaxies and lies about 0.2 
dex above that of the full galaxy sample, the dot-dashed line stems from blue galaxies 
lying about 0.15 dex below. The slope of the power law is about $\gamma \cong 1.4$ for 
all samples. For the remaining datasets we get similar results, however, the difference 
of the clustering strength between red and blue galaxies gets smaller as magnitudes increase.

The right panel of Fig. 2 shows the ratio between the full correlation functions of the 
sample m4 and all four mock catalogues. For clarity, error bars are only given for the upper 
and lower curves. The correlation functions of models C06 (solid line) and G11 (dot-dashed 
line) reproduce the shape of the observed correlation function over almost all spatial 
scales. However, the clustering amplitude is underpredicted by about 20 percent. Acceptable 
results are also obtained for the model D07, while F08 overpredicts the clustering of close 
pairs by up to a factor of two. The correlation function of other samples behave in a 
similar way.

\begin{figure}[!tH]
\vbox{
\centerline{\psfig{figure=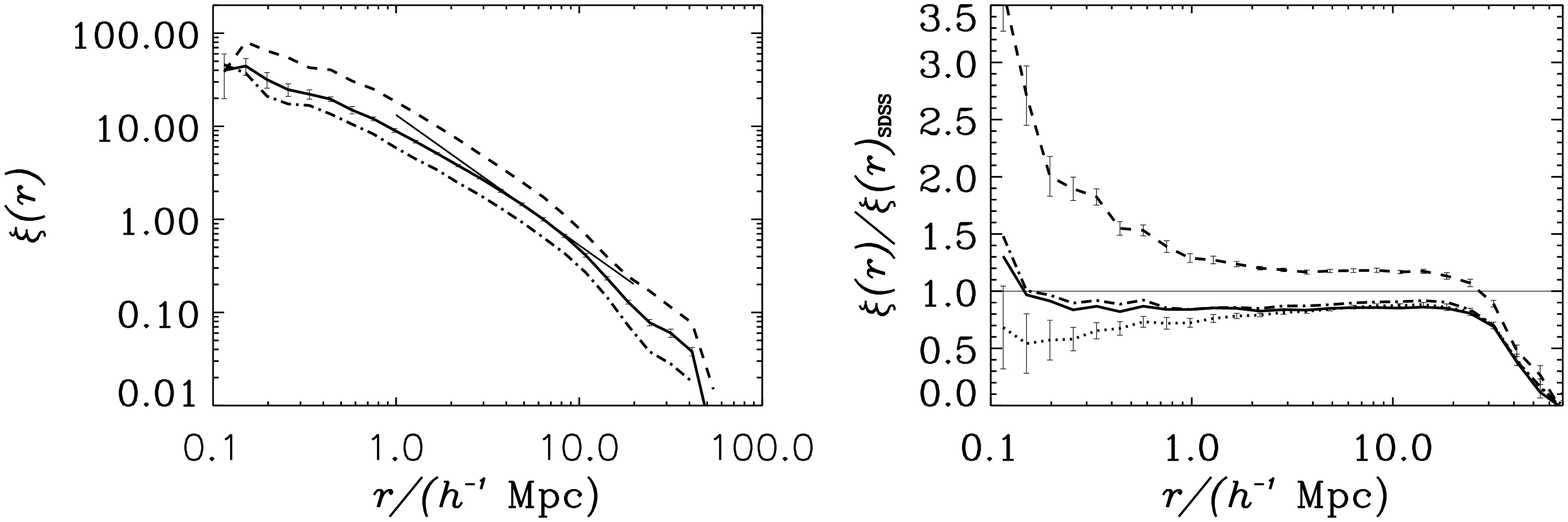,width=130mm,angle=0,clip=}}
\vspace{1mm}
\captionb{2}
{{\sl Left:} Two-point auto-correlation function for sample m1 with all galaxies 
(solid line), red galaxies (dashed line) and blue galaxies (dot-dashed line). For the full sample, 
a power law fit, $\xi = (r/r_0)^{1.4}$, centered at the correlation length scale, $r_0$, is shown. 
Error bars for the full sample are partly smaller than the line thickness. {\sl Right:} Ratio 
between model correlation functions and SDSS galaxies for the m4 sample. The different 
semi-analytic mock samples considered are those of C06 (solid line), D07 (dotted line with error 
bars), F08 (dashed line with error bars), and G11 (dash-dotted line).}
}
\end{figure}

The results can be described in a compact form evaluating the change of the correlation length 
as a function of absolute magnitude. The left panel of Fig.~3 shows the correlation length for 
the mean absolute $R$-magnitudes of samples m1 to m12. The solid, dashed and dot-dashed lines 
correspond to all, red, and blue galaxies, respectively. The correlation length difference 
between red and blue galaxies decreases from about 4 $h^{-1}$ Mpc at $R=-18.4$ to 2 $h^{-1}$ Mpc 
at $R=-21.5$. As seen in the figure, the brighter samples are dominated by red galaxies. The 
right panel shows the results corresponding to the G11 model. The correlation lengths of all and 
blue galaxies stay nearly constant between $R=-18.4$ and $R=-21$, while the correlation length of 
red galaxies decreases. This is due to the large number of satellites present among faint galaxies 
(cp. also Weinmann et al. 2006) that tend to cluster more strongly than field red galaxies with 
$R \cong -21$. At brighter magnitudes the correlation length increases due to the higher bias of 
more massive haloes with respect to the underlying mass distribution. The remaining semi-analytical 
models display similar trends. 

The ratio between the observed correlation length of red and blue galaxies and those corresponding 
to the semi-analytical models considered here can be seen in Fig. 4 (left and right panels respectively). 
In general, most models can explain the clustering amplitude of galaxies as measured by the correlation 
length with about 20 percent accuracy. However, there is a general trend for bright blue galaxies to be 
too weakly clustered. This is probably due to the fact that massive haloes display a too efficient star 
formation which therefore appear too bright for a given clustering strength. The trend showed by red 
galaxies is in principle similar. A remarkable exception can be seen at the faintest magnitude bin due 
to the efficient feedback implemented in the models. The other important exception is the increase 
observed for $R \lesssim -21$ in model C06  which is due to the strong quasar feedback implemented that 
makes bright red galaxies to be hosted by too massive and, therefore, too strongly clustered haloes.

\begin{figure}[!tH]
\vbox{
\centerline{\psfig{figure=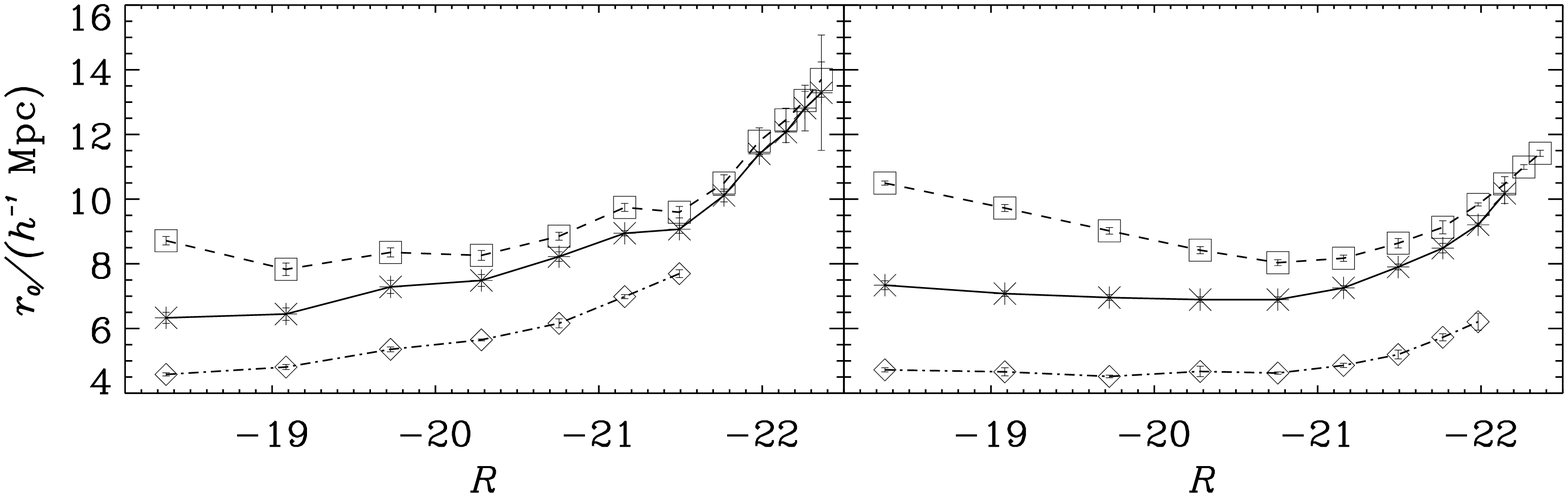,width=130mm,angle=0,clip=}}
\vspace{1mm}
\captionb{3}
{Correlation length as a function of mean $R$-magnitude. {\sl Left:} SDSS samples from m1 to m12 for all 
galaxies (stars and solid line), red galaxies (open squares and dashed lines), and blue galaxies (open 
diamonds and dash-dotted line). {\sl Right:} idem as left panel but for mock samples in the G11 model. 
Error bars represent 2 standard deviations, they are smaller than the symbols.}
}
\end{figure}

\begin{figure}[!tH]
\vbox{
\centerline{\psfig{figure=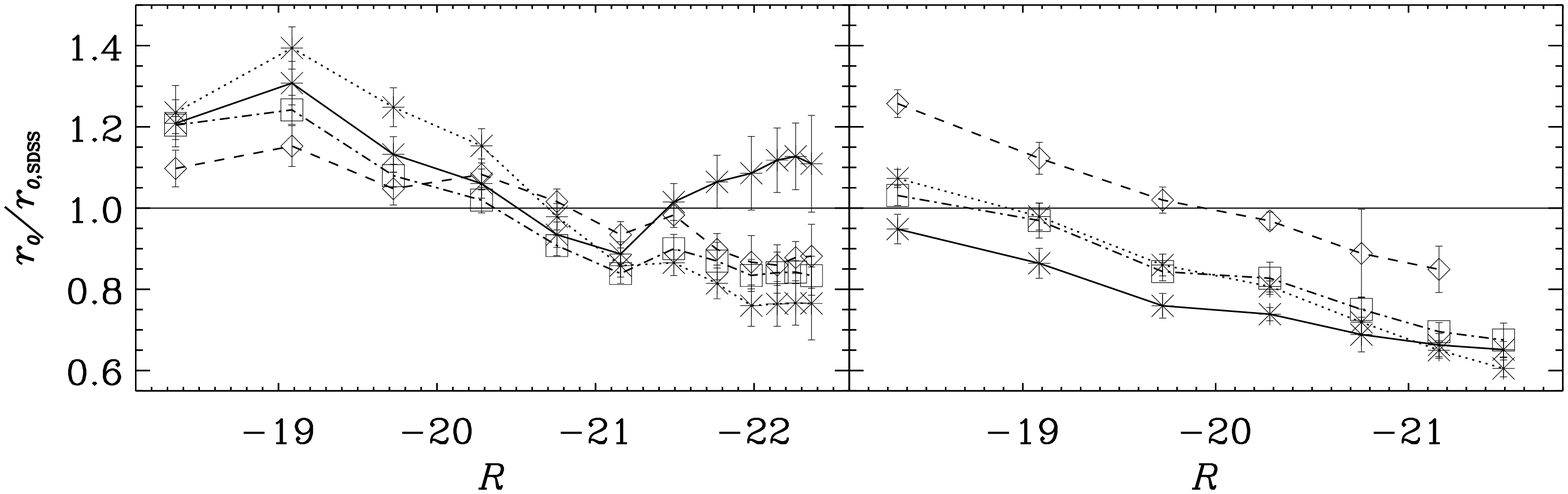,width=130mm,angle=0,clip=}}
\vspace{1mm}
\captionb{4}
{Ratio of the correlation length of mock and SDSS data as a function of mean $R$-magnitude. {\sl Left:} 
m1 to m12 samples for red galaxies. {\sl Right:} m1 to m7 samples for blue galaxies. The different 
semi-analytic mock samples considered are those of C06 (asterisks and solid line), D07 (asterisks and 
dotted lines), F08 (open diamonds and dashed line) and G11 (open squares and dot-dashed line). Errors
are again 2 standard deviations.}
}
\end{figure}

\sectionb{4}{MARK CORRELATION FUNCTION}

The trends already discussed for the clustering amplitude of galaxies using the standard two-point 
correlation function can be further investigated by means of the mark correlation function (e.g. Beisbart, 
Kerscher \& Mecke 2002). This statistical estimator is defined as the average of the inner galaxy properties 
$m$ -- here taken as color index $U-R$ or $R$ magnitude -- as a function of separation $r$ and can be written 
as ($\langle m \rangle$ is the average over the mark on the whole sample)

$$ k_m(r=|r_1-r_2|) = \frac{\langle m(r_1)+m(r_2)\rangle}{2\langle m\rangle}.$$

\noindent The left panel of Fig. 5 shows the mark correlation function of the samples m1 and m6 (solid 
and dashed lines respectively) compared to the corresponding mock samples for model F08 (dotted lines) 
using $U-R$ colours as a mark. Interestingly, there is a significant signal over a distance of about 
10 $h^{-1}$ Mpc where the samples show redder $U-R$ colours than the average. For the smaller scales this 
enhancement is about 0.05 to 0.1 mag. The excess of red neighbours is the result of the morphological 
transformation of galaxies by direct and tidal interactions. Since this effect is much stronger for faint 
galaxies it is natural to find a higher signal for sample m1. Below $1$ $h^{-1}$ Mpc our mock galaxies 
show a too strong mark correlation function. Obviously, the suppression of star formation in close galaxy 
pairs is overestimated in the models. The same behaviour is seen for the other mock samples.

As can be seen in the right panel of Fig. 5 when using absolute magnitudes as a mark the resulting signals 
are much weaker. The correlations for the samples r1 and r3 are shown as solid and dashed lines, while 
measures below and above $k_{U,R}=1$ correspond to $U$- and $R$-bands, respectively. This means that close 
pairs with a separation up to 10 $h^{-1}$ Mpc are brighter in the $R$ band and dimmer in the $U$ band by 
less than 0.005 mag. Despite the fact that the effect is weak, the result is significant as the corresponding 
error bars show. In this case errors are estimated using 100 samples with randomly reshuffled marks. 
 
\begin{figure}[!tH]
\vbox{
\centerline{\psfig{figure=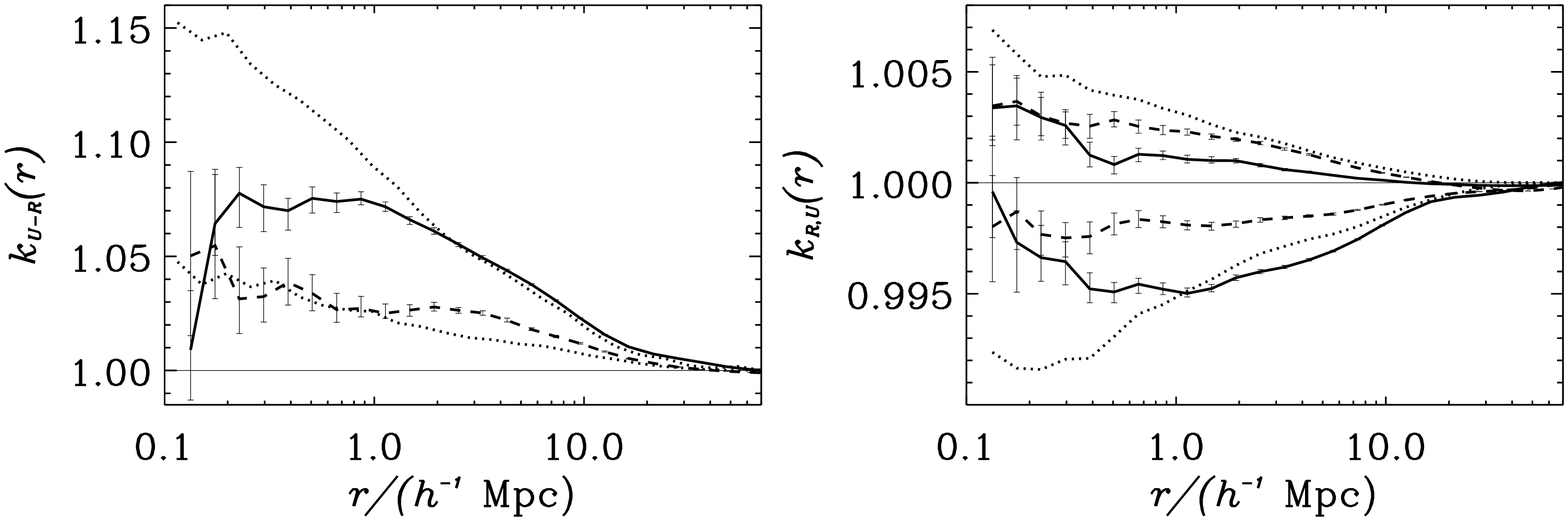,width=130mm,angle=0,clip=}}
\vspace{1mm}
\captionb{5}
{{\sl Left panel:} Mark correlation function using the $U-R$ colour as mark for samples m1 (solid line) and m6 
(dashed line) in comparison with mock samples given by the F08 model (dotted lines). {\sl Right:} Mark 
correlation function using the $U$-band ($k_U < 1$) and $R$-band ($k_R > 1$) absolute magnitudes as mark 
for samples r1 (solid line) and r3 (dashed line) in comparison with mock samples given by the F08 model 
(dotted lines). For clarity, mock galaxies here are only compared with sample r1. Errors are one standard
deviation.}
}
\end{figure}

\sectionb{3}{DISCUSSION}

The clustering of SDSS galaxies was previously discussed by Zehavi et al. (2010) mainly using the angular 
correlation function. Although this approach has the advantage of being independent of redshift space 
distortions, it uses only part of the information encoded in the galaxy distribution. However, results 
concerning the colour and magnitude dependence of clustering are similar to ours. Interestingly, the clustering 
of faint galaxies with $R \gtrsim -21$ is only weakly dependent on magnitude. In contrast, brighter galaxies 
are increasingly strong clustered as clearly seen from the luminosity dependence of the correlation function. 

We compared the clustering of SDSS galaxies with a large set of model galaxy samples based on the merger 
trees of the Millenium simulation that assume different semi-analytical prescriptions for galaxy formation 
models. These different theoretical models are able to qualitatively reproduce the clustering dependence 
as a function of magnitude and colour. However, quantitatively, still there exist significant differences, 
with the F08 model showing the smallest discrepancies for scales above  1 $h^{-1}$ Mpc. 

In addition to the standard two-point correlation technique, we carried out a new analysis using mark 
correlation functions which is suitable to assess the strength of galaxy transformations linked to their 
formation process. Surprisingly, we found a significant signal for galaxy pairs with a separation up to 
10 $h^{-1}$ Mpc depending on colour, and to a weaker extent, on absolute magnitudes.

It is our plan to continue the study of the properties of the galaxy distribution  and its connection with 
the large scale density field using mark correlation techniques. To characterize the density field we 
combine cosmological simulations with a galaxy group catalog to get the positions of suspected dark matter 
haloes. In extrapolating the mass density into the zones of influence of each halo we estimate the fine 
scale density field that reproduces both, the observed large scale galaxy distribution, and the average 
density profile around each group (Mu\~noz, M\"uller \& Forero-Romero 2011). This approach will therefore 
allow to further investigate the relation between the galaxy properties and their environmental density 
aiming at improving our knowledge of the cosmic web.

\thanks{VM is grateful to the organizers of the conference `Expanding the Universe' for the invitation and 
hospitality. SEN acknowledges support by the Deutsche Forschungsgemeinschaft under the grant MU1020 16-1.}

\References

\refb Beisbart C., Kerscher M., \& Mecke K.\ 2002, Lecture Notes in 
Physics, 600, 358 

\refb Blanton M. R., Schlegel D. J., Strauss M. A. et al. 2005, 
ApJ, 129, 2562


\refb Croton D. J., Springel V., White S. D. M. et al. 2006, 
MNRAS, 365, 11

\refb Davis M., Geller M. J. 1976, ApJ, 208, 13

\refb Davis M., Peebles P. J. E. 1983, ApJ, 267, 465

\refb De Lucia G., Blaizot J. 2007, MNRAS, 375, 2

\refb Doroshkevich A.~G., Saar E.~M., \& Shandarin S.~F.\ 1978, 
Proceedings of the IAU Symposium: Large Scale Structures in the Universe, 79, 423 

\refb Efstathiou G.\ 1993, Proceedings of the National Academy of Science, 90, 4859 

\refb Efstathiou G., Jedrzejewski R. I. 1984, Adv. Space Res., 3, 379

\refb Einasto J., Joeveer M. Saar, E. 1980, MNRAS, 193, 353

\refb Font A. S., Bower R. G., McCarthy I. G. et al. 2008, 
MNRAS, 389, 1619

\refb Geller M. J., Huchra J. P. 1989, Science, 246, 897

\refb Guo Q., White S. Boylan-Kolchin M. et al. 2011,
MNRAS 413, 101
 
\refb Huchra J., Davis M., Latham D., Tonry J. 1983, ApJS, 52, 89

\refb Li C., Kauffmann G., Jing Y. P. et al. 2006, MNRAS 368, 21

\refb Joeveer M., Einasto J., Tago E. 1978, MNRAS, 185, 357

\refb Longair M. S., Einasto J. 1978, Proceedings of the IAU 
Symposium: Large Scale Structures in the Universe, 79

\refb Loveday J., Maddox S. J., Efstathiou G., Peterson, B. A. 1995, 
ApJ, 442, 457

\refb Madgwick A. S., Hawkins E., Lahav O. 2003,
MNRAS 344, 847


\refb Mu\~noz J. C., M\"uller V., Forero-Romero J. E., 2011, MNRAS in press, arXiv:1107.1062

\refb Peebles P. J. E., Hauser M. G. 1974, ApJS, 28, 19

\refb Shane C. D., Wirtanen C. A. 1954, ApJ, 59, 285

\refb Springel V., White S. D. M., Jenkins A. et al. 2005, Nature, 435, 629

\refb Swanson M. E. C., Tegmark M., Hamilton A. J. S., Hill J. C. 2008,
MNRAS 387, 1391

\refb Totsuji H., Kihara T. 1969, PASJ 21, 221 

\refb Tucker D. L., Oemler Jr. A., Kirshner R. P. et al. 1997, 
MNRAS, 285, L5

\refb Weinmann Simone M., van den Bosch Frank C. et al. 2006, MNRAS 372, 1161

\refb Zehavi I., Zheng Z., Weinberg D. H. et al. 2010, arXiv 1005.2413

\refb Zel'dovich Ya. B. 1978, Proceedings of the IAU Symposium: Large Scale 
Structures in the Universe, 79, 409

\end{document}